\documentclass[draftclsnofoot, 12pt, onecolumn]{IEEEtran}  

\IEEEoverridecommandlockouts                              

\usepackage[utf8]{inputenc} 
\usepackage[T1]{fontenc}
\usepackage{url}
\usepackage{ifthen}
\usepackage{cite}

\usepackage{graphicx}
\usepackage{amssymb, amsmath, amsfonts}
\usepackage{tikz}
\usepackage{verbatim}
\usepackage{pifont}
\usepackage{float}
\usepackage{subfig}
\usepackage{caption}
\usetikzlibrary{patterns}
\newtheorem{theorem}{Theorem}
\newtheorem{corollary}[theorem]{Corollary}
\newtheorem{lemma}[theorem]{Lemma}

\newtheorem{defn}{Definition}

\newcommand{\diag}{\mathbf{diag}}

\newcommand{\suppress}[1]{}

\def\fb{\sf fb} 
\def\sbc{\sf sbc} 
\def\noisyfb{\sf noisy-fb} 
\def\nofb{\sf wo-fb} 

\def\sfb{\sigma_{\textup{fb}^2}}
\newcommand{\sfone}{\sigma_{\textup{fb}1}^{2}}
\newcommand{\sftwo}{\sigma_{\textup{fb}2}^{2}}
\newcommand{\stwo}{\sigma_{2}^2} 
\newcommand{\sone}{\sigma_{1}^2} 
\def\hs{\hat{\sigma}^2} 
\def\sfb{\sigma_{\fb}^2} 
\def\eE{\mathbb E} 

\def\Cnofb{\mathcal C^{\nofb}}
\def\Cwnfb{\mathcal C^{\noisyfb}}
\def\Cwefb{\mathcal C^{{\sf weak-fb}}}
\def\Caug{\mathcal{C}_{\sf{aug}}^{\noisyfb}}
\def\Caugnofb{\mathcal{C}_{\sf{aug}}^{\sf{wo-fb}}}

\def\thp{\theta^{\prime}} 
\def\independent{{\perp\!\!\!\perp}}

\def\Zfa{Z_{\sf fb1}}
\def\Zfb{Z_{\sf fb2}}

\def\Yta{\tilde Y_1}
\def\Ytb{\tilde Y_2}
\def\Yha{\hat Y_1}
\def\Yhb{\hat Y_2}
\def\thpp{\theta^{\prime\prime}}

\DeclareMathOperator{\Var}{Var}
\begin{document}
%
\title{On the Capacity Enlargement of Gaussian Broadcast Channels with Passive Noisy Feedback}
%
%
%
\author{Aditya~Narayan~Ravi,
        Sibi~Raj~B.~Pillai,
        Vinod Prabhakaran,
        Mich\`ele Wigger
\thanks{Parts of this paper were presented  at the Information Theory Workshop (ITW~2015)  and the International Symposium on Information Theory (ISIT~2020).}%
\thanks{The work of Mich\`ele Wigger has been supported by the ERC under grant agreement 715111. Vinod Prabhakaran acknowledges support of the Department of Atomic Energy, Government  of India, under project no. RTI4001. S.~R.~B.~Pillai acknowledges support from DST India under project EMR/2016/005847.
}%
\thanks{A.~Narayan~Ravi and S.~R.~B.~Pillai are with the Electrical Engineering Department, IIT Bombay, Mumbai, India. email:\{adityan,bsraj\}@ee.iitb.ac.in}%
\thanks{V.~M.~Prabhakaran is with the School of Technology and Computer Science, TIFR Mumbai, India. email:vinodmp@tifr.res.in}%
\thanks{M.~Wigger is with LTCI Telecom Paris, IP Paris, France. email:
                     michele.wigger@telecom-paristech.fr.}
}

\maketitle

\begin{abstract}
It is well known that the capacity region of an average transmit power constrained Gaussian Broadcast Channel (GBC) with 
independent noise realizations at the receivers is enlarged by the presence of causal noiseless feedback. Capacity region
enlargement is also known to be possible by using only passive \emph{noisy} feedback, when the GBC has identical noise  
variances at the receivers. The last fact remains true even when the feedback noise variance is very high, and available only from 
one of the  receivers. While such capacity enlargements are feasible for several other feedback 
models in the Gaussian BC setting, it is also known that feedback does not change the capacity region for physically degraded broadcast channels.
In this paper, we consider a two user GBC with independent noise realizations at the receivers, where the
feedback links from the receivers are corrupted by 
independent additive Gaussian noise processes. We investigate the set of four noise variances, two forward and two feedback, for which no capacity enlargement 
is possible. A sharp characterization of this region is derived, i.e., any quadruple outside the presented region will 
lead to a capacity enlargement, whereas quadruples inside will leave the capacity region unchanged. Our results lead to the 
conclusion that when the forward noise variances are different, too noisy a feedback from one of the receivers alone is
not always beneficial for enlarging 
the capacity region, be it from the stronger user or the weaker one, in sharp contrast to the case of equal forward 
noise variances.


\end{abstract}


%
\IEEEpeerreviewmaketitle

\section{Introduction}

\IEEEPARstart{F}{eedback} does not increase the capacity of a memoryless
point-to-point channel, a result which goes back to C.~E.~Shannon
\cite{Shannon1956zero}. However, feedback has a positive impact in simplifying coding
schemes and boosting error exponents~\cite{GamalKim11}. With the discovery of
capacity regions for several multiuser models in the '70s and '80s, it was of significant
interest to find the impact of feedback on these models. In the absence of feedback, identifying suitable
auxiliary variables which can lead to  single letter expressions for the rate-region
turned out to be a key step in discovering the capacity region of a degraded Broadcast
Channel (BC)~\cite{Gallager74}. 
Notice that, without any feedback, one can turn a  stochastically degraded BC to 
an equivalent physical degraded setup, as only the marginal 
distributions  to the individual receivers matter in decay of the error probability
with blocklength. 
When perfect causal feedback is available in a two user BC, 
a single letter characterization of the capacity region in terms of one auxiliary variable 
is obtained by El Gamal in~\cite{Gamal78a} and~\cite{Gamal81b}, where 
the physical degradation assumption is crucial. More specifically,~\cite{Gamal78a} and~\cite{Gamal81b}
respectively show that the capacity regions of
a physically degraded BC and a physically degraded Gaussian BC (GBC) are unchanged 
by the presence of causal feedback.
%
In contrast, Dueck~\cite{Dueck80} demonstrated a BC for which rate pairs
outside its no-feedback capacity region can be attained using feedback. 
For a  stochastically degraded two user GBC, Ozarow and Leung developed a feedback coding scheme to show that perfect noiseless feedback from both the receivers enlarges the  capacity region~\cite{ozarow1984}, when there is no physical degradation. It was later shown that perfect noiseless feedback from the stronger receiver was sufficient to enlarge the capacity region~\cite{bhaskaran2008}.
The optimism of capacity enlargement using feedback did carry over to a variety of models. Recent works considered BCs  with noisy feedback~\cite{shayevitz2012},~\cite{venkataramanan2013},~\cite{Love15} and rate limited feedback~\cite{wu2016}. In fact, for a two user GBC with equal receiver noise variances, passive noisy feedback from any one of the receivers enlarges the capacity region, even when the feedback noise is of very high variance~\cite{venkataramanan2013}. 

It was  shown in \cite{lapidoth2010} that noisy feedback always enlarges the capacity region of a Gaussian Multiple Access Channel (MAC), a fact which remains true with the  availability of feedback to only one of the transmitters. Furthermore, a MAC-BC duality  while employing linear feedback coding schemes with noiseless feedback is known~\cite{amor2014mac}, allowing many of the MAC results to be relevant for the BC as well. In another related result, \cite{wigger2008pre} showed that the gains due to feedback can be potentially unbounded in correlated noise channels. In summary,  capacity enlargement for Gaussian BCs using noisy feedback turned out to be true for several models which are not physically degraded. That this is not always the case when the receivers have different noise variances is shown in the present paper. The results here significantly expand some of the initial results in the conference versions~\cite{PilPra2015, Aditya2020}, which considered one sided feedback from the stronger receiver.  The main interest here is in identifying the set of four noise variances, corresponding to two forward noise processes and two feedback noise processes, such that the capacity region differs from that without feedback. We wish to highlight the following aspects of the paper:
\begin{itemize}
    \item noisy feedback from both the receivers are considered.
    \item the exact threshold behaviour is characterized, i.e., any set of four noise variances can be classified based on whether the capacity region of the GBC with feedback, having these parameters, is enlarged or not.
    \item that too noisy a feedback from the weaker receiver of a two user GBC does not enlarge the capacity region was hitherto unknown.
\end{itemize}
\subsection{Notations:}
For a positive integer $n$, we use $U^n$  to denote the tuple $U_1,U_2, \cdots,U_n$. The diagonal square matrix of size $n \times n$ is denoted by  $\diag(\bar d)$, where the diagonal elements are given by the vector $\bar d$. We write $U \sim \mathcal{N}(\mu,\mathbf{K})$ to denote a random vector $U$ having a Multivariate Gaussian Distribution with covariance matrix $\mathbf{K}$. The
acronyms LHS and RHS stand for Left Hand Side and Right Hand Side respectively, of the  mathematical equation in consideration.  Logarithms are taken with respect to base $2$.
 
\subsection{Paper Organization:}
The organization of the paper is as follows. We introduce the two user scalar GBC with passive noisy feedback from both the receivers in the next section, and present our objectives and the main result.
Before we proceed to the detailed proof, we take a detour in Section~\ref{sec:useless} to present results on the utility of feedback in some related channel models.
Then, in order to prove the main result,
we first develop a converse argument in Section~\ref{sec:converse:both}, to show there are 
regimes of possible noise variances (four parameters, two forward and two feedback) 
where  the capacity region remains the same as that without feedback. The boundary of the regime thus characterized is then shown to be sharp, by constructing an achievable region which strictly enlarges the no feedback capacity for noise parameters exterior to the regime.
This is presented in Section~\ref{sec:achieve}. Finally, Section~\ref{sec:conc}
concludes the paper.

\section{Model and Results}\label{sec:model}
\begin{figure}[htbp]
\begin{center}
\begin{tikzpicture}[line width=1.5pt]
\coordinate (cenc) at (0,0);
\coordinate  (cdec1) at  (4.75,1.25);
\coordinate  (cdec2) at  (4.75,-1.25);
\coordinate  (cz1) at  (2.75,1.25);
\coordinate  (cz2) at  (2.75,-1.25);

\node (enc) at (cenc)[rectangle, draw, text width=1.2cm, minimum height=1.25cm]{Sender};
\node (z1)  at (cz1)[circle, draw] {$+$};
\node (z2)  at (cz2)[circle, draw] {$+$};
\node (dec1) at (cdec1)[rectangle, draw, text width=1cm, minimum height=1.0cm]{Rec~1};
\node (dec2) at (cdec2)[rectangle, draw, text width=1cm, minimum height=1.0cm]{Rec~2};
\draw[->] (z1) -- node (cfb1) [pos=0.5]{} (dec1); \draw[->] (z2) -- node (cfb2) [pos=0.5]{} (dec2);
\draw[<-] (enc) --++(-1.05,0) node[left]{$W_1, W_2$};
\draw[->] (enc) -- node[above, pos=0.4]{$X_i$} ++(1.5,0) |- (z1);
\draw[->] (enc) -- ++(1.5,0) |- (z2);
\path (cfb1) --++(0,1) --++(-2.75,0) node (zfb1)[circle, draw]{$+$};
\path (cfb2) --++(0,-1) --++(-2.75,0) node (zfb2)[circle, draw]{$+$};
\draw[->] (cfb1.center) |- (zfb1);
\draw[->] (zfb1) -| (enc);
\draw[->] (cfb2.center) |- (zfb2);
\draw[->] (zfb2) -| (enc);
\draw[->] (dec1) --++(1,0) node[right, xshift=-0.1cm]{$\hat W_1$};
\draw[->] (dec2) --++(1,0) node[right,xshift=-0.1cm]{$\hat W_2$};
\draw[<-] (z1) --++(0,-0.7)node[below]{$Z_{1,i}$};
\draw[<-] (z2) --++(0,0.7)node[above]{$Z_{2,i}$};
\draw[<-] (zfb1) --++(0,-0.75)node[below]{$Z_{\text{fb1},i}$};
\draw[<-] (zfb2) --++(0,0.75)node[above]{$Z_{\text{fb2},i}$};
\end{tikzpicture}
\caption{Scalar GBC  with passive noisy feedback from both receivers.} \label{fig:bc:one}
\end{center}
\end{figure}
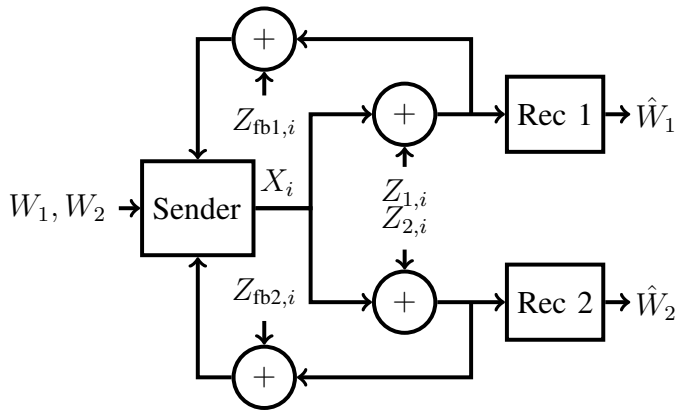

Consider a memoryless two user scalar GBC, as shown in Fig.~\ref{fig:bc:one}. Assume independent memoryless
noisy feedback links from both receivers to the transmitter.
In this setup, $X$ refers to the signal transmitted by the sender, and the additive noise processes $(Z_1,Z_2)$ in the forward links are distributed according to $\mathcal N(0,\text{diag}(\sone, \stwo))$. Unless otherwise stated, we will assume
\begin{equation}
    \sone \leq \stwo,
\end{equation}i.e., receiver 2's outputs are  more noisy than receiver 1's. The passive causal feedback links are corrupted by additive noise $(Z_{\text{fb}1},Z_{\text{fb}2}) \sim \mathcal N(0,\text{diag}(\sfone, \sftwo))$,
independent of the forward noise processes. 

A pair of messages $(W_1,W_2)$, independently and uniformly chosen from $\{1,\cdots, 2^{nR_1}\} \times \{1,\cdots, 2^{nR_2}\}$ is to be conveyed in $n$ channel uses of the GBC. 
The encoder chooses the symbol $X_i$ at time $i\in\{1,\ldots,n\}$ as a function of both the messages as well as causal noisy feedback from both receivers:
\begin{align}
    X_i = g_i(W_1,W_2,Y_1^{i-1} + Z_{\text{fb}1}^{i-1},Y_2^{i-1} + Z_{\text{fb}2}^{i-1}).
\end{align}
The transmissions are constrained to meet an
average power constraint $P$, i.e. $\sum_{i=1}^n \eE |X_i|^2 \leq nP$. After every block of $n$ transmissions, each receiver attempts to decode the message intended to it. Let $P_e(n)$ denote the average error probability that any of the two receivers makes a  decoding error.
 We are interested in the capacity region $\Cwnfb$ of this model. Using standard definitions~\cite{GamalKim11}, the capacity region can be concisely defined as follows.
\begin{defn}
The capacity region $\mathcal C_{\sbc}^{\noisyfb}$ is the convex closure of the set of
all nonnegative rate pairs $(R_1,R_2)$
	such that there exists a sequence of encoder-decoder pairs with $P_e(n)
\rightarrow 0$ as $n\rightarrow \infty$.
\end{defn}

Let $\Cnofb$ refer to the capacity region when there are no feedback links present, well known to be achieved by superposition coding~\cite{CovTho91}. 
$\Cnofb$ is the collection of nonnegative rate pairs $(R_1,R_2)$ such that
\begin{subequations}\label{eq:nofb}
\begin{IEEEeqnarray}{rCl} 
    R_1 &\leq &\frac{1}{2}\log\left(1 + \frac{\theta P}{\sone}\right)  \\
    R_2 & \leq &\frac{1}{2}\log\left(1 + \frac{(1-\theta )P}{\theta P + \stwo}\right),
\end{IEEEeqnarray}
\end{subequations}
for some $\theta \in [0,1]$. By ignoring feedback, it is easy to see
\begin{align} \label{eq:inc}
    \Cnofb \subseteq \Cwnfb.
\end{align}
For equal noise variances, i.e. $\sigma_1^2=\sigma_2^2$, Venkatramanan and Pradhan~\cite{venkataramanan2013} showed that the inclusion in \eqref{eq:inc} is  strict, irrespective of the feedback noise variances. 
In fact, it is shown that noisy feedback from even one of the two receivers always enlarges the capacity region. This brings the following question
to the fore: For what set of parameters $(\sone, \stwo, \sfone, \sftwo)$ in $(\mathbb{R^+} \cup   \{\infty\})^4$ will the inclusion in
$\eqref{eq:inc}$ stays strict? Alternately, are there regimes of noise variances where feedback is rendered futile in enlarging the capacity region? Our main result is the following.
\begin{theorem} \label{thm:main}
For $\sone\leq \stwo$, the relation  $\Cnofb = \Cwnfb$ holds if and only if  
\begin{align} \label{eq:thm:main}
  \frac{\stwo}{\sone} \geq \left(\frac{\sone}{\sfone} + 1\right)\left(\frac{\stwo}{\sftwo} + 1\right).
\end{align}
\end{theorem}
\begin{IEEEproof}
The proof involves two main steps. One is a converse argument to show that the capacity region is unchanged if the condition in \eqref{eq:thm:main} is met. The second part
requires an achievable communication scheme, which operates  at rates outside
the no feedback capacity region when \eqref{eq:thm:main} is not met.
These details will be provided in the next two sections.
\end{IEEEproof}

The following corollaries can be obtained by simple algebraic manipulations.

\begin{corollary}[One-sided feedback]\label{cor:Onesided}
Consider a GBC with $\sone\leq \stwo$. When  feedback is available only from receiver~$2$, i.e.  $\sfone=\infty$,
the relation $\Cnofb = \Cwnfb$ holds if and only if
\begin{equation}
 \sftwo \geq\stwo \cdot \left(\frac{\stwo}{\sone}-1\right)^{-1}.
\end{equation}
On the other hand, if feedback is only available from receiver~$1$, i.e. $\sftwo=\infty$, 
then the relation $\Cnofb = \Cwnfb$ holds if and only if
\begin{equation}
    \sfone \geq \sone \cdot \left(\frac{\stwo}{\sone}-1\right)^{-1}.
\end{equation}
\end{corollary}

Above corollary can be rephrased to show that one-sided feedback from receiver 1 increases capacity if and only if $\frac{\stwo}{\sfone} \leq\frac{\sone}{\sfone}\left( \frac{\sone}{\sfone}+1\right)$. Similarly, one-sided feedback from receiver 2 increases capacity if and only if $\frac{\sone}{\sftwo} \geq\frac{ \stwo/\sftwo}{\stwo/\sftwo +1}$. 

Fig.~\ref{fig:sbc-plot} depicts the regimes of  noise-variances where capacity is enlarged with one-sided feedback from receiver 1. The figure also allows for $\sone > \stwo$. In this case, the desired result  is obtained by swapping  indices $1$ and $2$ in above corollary.  The figure illustrates that  for $\sone=\stwo$, one-sided feedback from receiver~1 is always helpful, no matter how noisy. Otherwise, if $\sone \neq \stwo$, then one-sided feedback from receiver~1 becomes useless when the feedback noise-variance exceeds a certain threshold. Similar statements hold also for one-sided feedback from receiver 2. In fact, the corresponding plot can be obtained from Fig.~\ref{fig:sbc-plot} by mirroring the plot on the $\sone=\stwo$ line.

Theorem~\ref{thm:main} further provides the following corollary on the usefulness of feedback.
\begin{corollary}
If  $\sone=\stwo$, feedback enlarges the capacity region whenever $\sfone <\infty$ or $\sftwo < \infty$.

On the contrary, if $\sone \neq \stwo$  there are finite feedback noise variances $\sfone<\infty$ and $\sftwo<\infty$ such that even feedback from both users does not enlarge capacity. 
\end{corollary}
\begin{IEEEproof}The proof follows by noting that for $\sone=\stwo$, the LHS of \eqref{eq:thm:main} evaluates to 1 and is thus dominated by its RHS whenever $\sfone$ or $\sftwo$ are finite. For $\sone < \stwo$, the LHS of \eqref{eq:thm:main} exceeds 1 and it is possible to find finite $\sfone$ and $\sftwo$ such that the RHS of \eqref{eq:thm:main} does not exceed its LHS.
\end{IEEEproof}
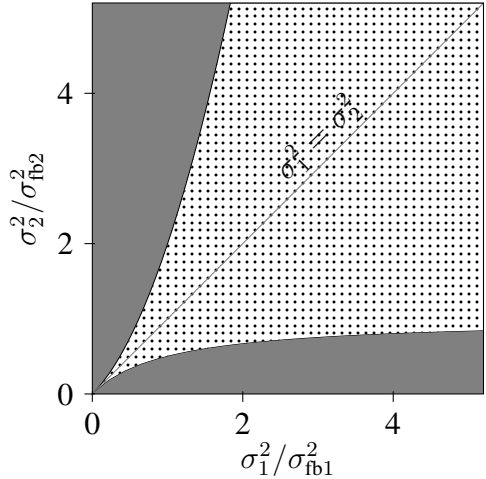
\begin{figure}
\begin{center}
\usetikzlibrary{decorations.markings,decorations.shapes}
\begin{tikzpicture}[label/.style={%
   postaction={ decorate,transform shape,
   decoration={ markings, mark=at position .5 with \node #1;}}}]
    \fill[fill=gray] (0,0) -- (0,5.2) -- plot [domain={(sqrt(1+4*5.2)-1)/2}:0] (\x,{\x+\x^2})  -- cycle;
   \draw[gray!50,pattern = dots] plot [domain=0:{(sqrt(1+4*5.2)-1)/2}] (\x,{\x+\x^2}) --({(sqrt(1+4*5.2)-1)/2},5.2) -- (5.2,5.2) --  plot [domain=0:5.2] (\x,5.2)  -- plot [domain = {5.2}:0] (\x,{\x/(\x + 1)}) -- cycle;
    \draw plot[domain=0:{(sqrt(1+4*5.2)-1)/2}] (\x,{\x+\x^2});
    \draw plot[domain=0:5.2] (\x,{\x/(\x + 1)});
  \fill[fill=gray] plot[domain=0:5.2] (\x,{\x/(\x + 1)}) -- plot [domain=5.2:0] (5.2,\x) -- plot [domain=0:5.2] (\x,0) -- cycle;
    \draw[thin, gray] plot[domain=0:5.2] (\x,\x) node[xshift=-2.2cm,
    yshift=-1.8cm, rotate=45, black, opacity=0.95]{$\sone=\stwo$};
        \draw [thin] (0,0) --  coordinate (xaxis) (5.2,0) -- (5.2,5.2) --
(0,5.2) -- coordinate (yaxis) (0,0);
        \foreach \x in {0,2,4}
                \draw (\x,1pt) -- (\x,-3pt)
                        node[anchor=north] {\x};
        \foreach \y in {0,2,4}
                \draw (1pt,\y) -- (-3pt,\y)
                        node[anchor=east] {\y};
        \node[below=0.5cm] at (xaxis) {${\sone}/{\sfone}$};
        \node[rotate=90,above=0.5cm] at (yaxis) {${\stwo}/{\sftwo}$};
\end{tikzpicture}
\end{center}
\caption{The patterned region indicates where one-sided feedback from receiver 1 enlarges capacity, and the shaded region indicates where capacity remains unchanged.} 
\label{fig:sbc-plot}
\end{figure}


\section{More On the Usefulness of Feedback}\label{sec:useless}
In this section we consider two related BC setups where the feedback links to the transmitter do not increase the capacity region. 
The results here may be of interest on their own. Moreover, Corollary~\ref{cor:Onesided} can be proved using the first result that we present in this section, stated as Theorem~\ref{thm:not_useful} below. However, the results and tools presented in this section  seem not enough to prove the converse to our more general result presented in  Theorem~\ref{thm:main}.

\subsection{A Vector-BC with Partial Feedback}\label{sec:partial}
We start with a slight extension of El Gamal's negative  result on the usefulness of feedback for physically degraded BCs, to vector channels with  partial feedback. Consider a general (not necessarily Gaussian) BC where the first receiver observes $Y_1=(\Yta, \Yha)$, and the second receiver 
observes $Y_2=(\Ytb, \Yhb)$. Let the transition law of the BC be described by $P_{\Yta,\Yha,\Ytb,\Yhb|X}$.  
Feedback is only from outputs $\Yta$ and $\Ytb$ to the transmitter, and can be noisy or perfect.
The following negative result holds.
\begin{theorem}\label{thm:not_useful}
Consider a BC of transition law $P_{\Yta,\Yha,\Ytb,\Yhb|X}$.
If there exists a conditional law  ${P}_{\Ytb,\Yhb|\Yta,\Yha}^{\prime}$ such that the joint law 
\begin{equation}\label{eq:tilde_law}
{P}_{\Yta,\Yha,\Ytb,\Yhb |X}^{\prime} \triangleq  {P}_{\Ytb,\Yhb|\Yta,\Yha}^{\prime} \cdot {P}_{\Yta\Yha|X}
\end{equation}
    satisfies 
    \begin{subequations}\label{eq:assumptions}
    \begin{IEEEeqnarray}{rCl}
        {P}_{\Yha|X,\Yta,\Ytb}^{\prime}  &= & P_{\Yha|X,\Yta,\Ytb} , \label{eq:a1}\\
	    {P}_{\Yhb|X,\Yta,\Ytb}^{\prime} &= & P_{\Yhb|X,\Yta,\Ytb},  \label{eq:a2}\\
	 {P}_{\Yta,\Ytb|X}^{\prime} & = & {P}_{\Yta,\Ytb|X}, \label{eq:a3}
        \end{IEEEeqnarray}
        \end{subequations}
    then
         causal feedback of the two outputs $\Yta$ and $\Ytb$ to the transmitter does not enlarge the capacity region. 
\end{theorem}
\begin{IEEEproof}
Consider the BC  
of transition law ${P}_{\Yta,\Yha,\Ytb,\Yhb |X}^{\prime}$ as defined in \eqref{eq:tilde_law}. This channel is indeed 
physically degraded because its law  satisfies the Markov chain $X \to (\Yta,\Yha) \to (\Ytb,\Yhb)$.
Moreover, by Assumptions~\eqref{eq:assumptions}, under any encoding strategy, and both in the presence and absence of feedback, the joint law 
	of the tuples $(W_1, \Yta^n,\Yha^n)$  and $(W_2, \Ytb^n,\Yhb^n)$ is the same in both the original BC $P_{\Yta,\Yha,\Ytb,\Yhb|X}$
	and the physically degraded version  $P_{\Yta,\Yha,\Ytb,\Yhb|X}^{\prime}$. Since the error probability at 
each receiver only depends on its own observations, but not on the observations at the other receiver, the capacity regions of the original and the physically degraded BCs coincide.

On any physically degraded BC, the capacity region, even  with full causal feedback, remains the same as that  without 
feedback~\cite{Gamal81b,Gamal78a}. By the  above observations, the same must also  hold for our original BC, thus concluding
	the proof.
\end{IEEEproof}

\subsection{Feedback from the Weak Receiver to all Terminals}\label{sec:fb_all}

\begin{figure}[H]
	\begin{center}
	\subfloat[Original BC~\label{fig:or1}]{
	\begin{tikzpicture}[line width=1.0pt]
	
	\node (n0) at (0,0) {$X$};
	\node (r1) at (3,1) {$Y_1$};
	\node (r2) at (3,-1) {$Y_2$};
	\draw[->] (n0) -- (r1);
	\draw[->] (n0) -- (r2);
	\draw[dashed,->] (r2) -- coordinate[pos=0.5] (cf)  ++(2,-0.66) node[right] (rc) {$Y_c$};
	\node (zf) at (cf)[fill=white]  {$+$};
	\draw[<-] (zf) --++(-0,0.5) node[above]{$\Zfb$};
	\draw[fill=white] (cf) circle(0.25cm) node {$+$};
	\draw[dashed, ->] (rc) to[bend left=60] (n0);
	\draw[dashed, ->] (rc) to[bend right=30] (r1);
	\end{tikzpicture}} \hspace*{2.0cm}
	\subfloat[Augmented BC~\label{fig:common:fb:new}]{
	\begin{tikzpicture}[line width=1.0pt]
	\node (n0) at (0,0) {$X$};
	\node (r11) at (4,1){$Y_1$};
	\node (r12) at (4,2){$Y_c$};
	\node (r21) at (4,-1){$Y_a$};
	\node (r22) at (4,-2){$Y_c$};
	\draw[->] (n0) -- (r11);
	\draw[->] (n0) -- (r12);
	\draw[->] (n0) -- (r21);
	\draw[->] (n0) -- (r22);
	\draw[dashed, ->] (r22.210) to[bend left=50] (n0);
	\end{tikzpicture}
	}
	\caption{Original and Augmented BCs}
	\end{center}
	\end{figure}
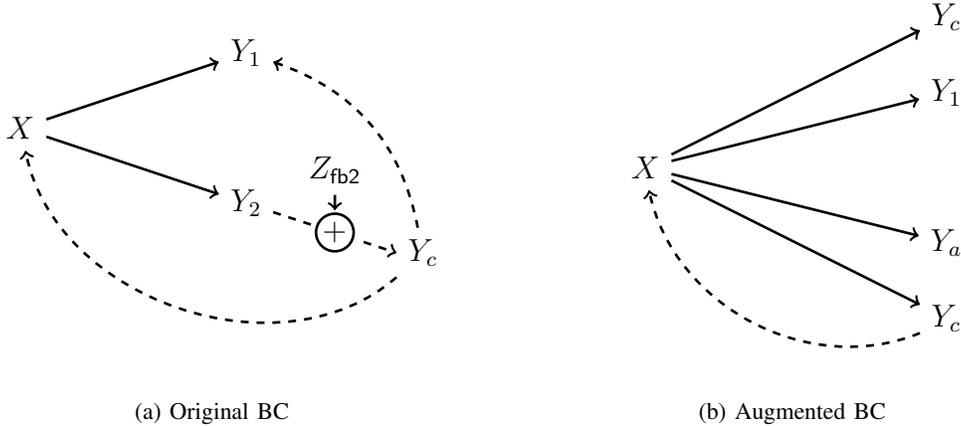

Recall the stochastically degraded GBC in Section~\ref{sec:model}, where in particular $\sone \leq \stwo$. 
Let us assume one-sided noisy feedback from receiver~2 to the transmitter. In addition,
suppose that the same feedback values are 
observed at receiver~$1$ as well.
So, receiver~1 observes its own channel outputs $Y_1^n$ as well as the noisy feedback outputs 
	\begin{equation}\label{eq:Ycn}
	Y_{\textnormal{c}}^n \triangleq Y_2^n+Z_{\fb2}^n, 
	\end{equation}
	which are causally  observed also at the transmitter. The model is illustrated  in Fig.~\ref{fig:or1}. We denote  its capacity region by $\Cwefb$. 
Notice that the described setup is  physically degraded only if the feedback noise $Z_{\fb2}^n$ is deterministic and thus $\sftwo=0$. Interestingly, Ozarow considered this degraded setup in~\cite{ozarow1984} to derive an outer bound to the GBC capacity region with perfect feedback.

The following theorem is the main result of this subsection and determines the feedback capacity region $\Cwefb$. Interestingly, the result shows that the feedback link \emph{to the
transmitter} has no effect on the capacity region  of the described model. In other words, once receiver~$1$ learns the feedback signal from receiver~$2$, the capacity region does not depend on whether also the transmitter observes the feedback or not. 
\begin{theorem} The capacity region $\Cwefb$ is the same as that without feedback to the transmitter, and is characterized as:
	\begin{equation}\label{eq:Cwefb}
	\Cwefb = \bigcup_{\theta \in [0,1]} \left\{ (R_1, R_2): 
		R_1 \leq \log\left(1+ \frac{\theta P}{{\sigma}_{\textnormal{comb}}^2}\right)\text{ and }
		R_2 \leq \log\left(1+\frac{(1-\theta)P}{\theta P + \stwo}\right)
		\right\},
	\end{equation}
	where 
	\begin{equation}\label{eq:sigma_comb}
	{\sigma}_{\textnormal{comb}}^2 \triangleq  \frac{\sone(\stwo+\sftwo)}{\sone+\stwo+\sftwo}.
	\end{equation}
\end{theorem}
\begin{IEEEproof}
Since $\Cwefb$ is same as the capacity region without feedback, its
achievability follows from superposition coding and maximum ratio combining at receiver~$1$.

The key steps in the converse are to  show that the capacity region $\Cwefb$ is included in the capacity region of the \emph{augmented} BC in Fig.~\ref{fig:common:fb:new}, and  to invoke Theorem~\ref{thm:not_useful} to show that feedback does not increase the capacity of this augmented BC. The final step is then to argue that the capacity region without feedback of the augmented BC coincides with the RHS of \eqref{eq:Cwefb}. 

Consider the augmented BC depicted in Fig.~\ref{fig:common:fb:new}. Receiver~1 observes outputs $(Y_1^n, Y_c^n)$ and receiver~2 observes outputs $(Y_a^n, Y_c^n)$, where
\begin{equation}\label{eq:Yan}
    Y_a^n \triangleq  Y_2^n - \alpha Z_{\fb2}^n= X^n +\underbrace{ Z_2^n -\alpha Z_{\fb2}^n}_{\triangleq Z_a^n}
\end{equation}
and 
	\begin{equation} \label{eq:sec3:alpha}
    \alpha \triangleq \frac{\stwo}{\sftwo}. 
\end{equation}
The transmitter observes the feedback output $Y_{c,i}$ after each channel use $i$. The capacity region $\Caug$ of this augmented BC with noisy feedback includes the original capacity region $\Cwefb$, i.e.
\begin{equation}\label{eq:lin1}
    \Cwefb \subseteq \Caug
\end{equation}
because receiver~$1$ in the augmented BC observes the same outputs as in the original BC, and receiver~$2$ can compute  $Y_2^n$ from its observations $(Y_c^n, Y_a^n)$, see \eqref{eq:Ycn} and \eqref{eq:Yan}.

Notice next that feedback does not increase the capacity of the augmented BC, because this BC satisfies the assumptions in Theorem~\ref{thm:not_useful}. 
To see this, specialize the theorem to $\Yta = Y_c$, $\Yha = Y_1$, $\Ytb=Y_c$, and $\Yhb = Y_a$.
Now  set  $P_{\Ytb|\Yha,\Yta}^{\prime}=P_{\Ytb|\Yha,\Yta}=P_{\Ytb|\Yta}$ (i.e. $\Yta=\Ytb$ under both $P$ and $P^{\prime}$), 
and choose ${P}_{\Yhb|\Yha,\Yta,\Ytb}'=P_{\Yhb|\Yha}^{\prime}$ in such a way that 
\begin{equation}
    \Yhb = \Yha + \hat{Z}_2
\end{equation}
for  $\hat{Z}_2$ a zero-mean Gaussian random variable of variance $\stwo + \alpha^2 \sftwo- \sone$ and independent of all other random variables. This is
possible since $\sigma_2^2 \geq \sigma_1^2$.

Under this choice, \eqref{eq:a1}  holds since $\hat Y_1 = X + Z_1$ under both $P$ and $P^{\prime}$,
 with $Z_1 \independent (Z_2,\Zfb)$.
The assumption \eqref{eq:a2} also holds, since the pair $(Z_a, Z_2+\Zfb)$ has independent Gaussian
entries and it has the same distribution as $(Z_1+\hat Z_2,Z_2 + \Zfb)$.
	Since $\Yta=\Ytb=Y_c$, the condition \eqref{eq:a3} holds as well. Thus, 
	we can employ Theorem~\ref{thm:not_useful} to conclude that feedback does not increase the capacity of the augmented BC, i.e.,
	\begin{equation} \label{eq:aug=noaug}
\Caug = \Caugnofb,
\end{equation}
where $\Caugnofb$ denotes the capacity region of the augmented BC without feedback.
The capacity region without feedback of the augmented BC is obtained from 
the  well-known  capacity region of a Gaussian vector BC: 
\begin{equation}\label{eq:lin2}
    \Caugnofb = \bigcup_{\theta \in [0,1]} \left\{ (R_1, R_2): 
		R_1 \leq \log\left(1+ \frac{\theta P}{{\sigma}_{\textnormal{comb}}^2}\right)\text{ and }
		R_2 \leq \log\left(1+\frac{(1-\theta)P}{\theta P + \stwo}\right)
		\right\},
\end{equation}
where $\sigma^2_{\textnormal{comb}}$ is defined in \eqref{eq:sigma_comb}. 
 Notice that here we used the fact that receiver 2 can compute $Y_2$ from the pair $(Y_a,Y_c)$ and that, due to the Markov chain $X \to Y_2 \to (Y_a,Y_c)$, the pair $(Y_a,Y_c)$ is redundant in view of $Y_2$.
 
Combining \eqref{eq:lin1}, \eqref{eq:aug=noaug} and \eqref{eq:lin2} establishes the desired converse proof. 
\end{IEEEproof}

 
	%

\section{Outer Bounds for GBC with Noisy Feedback (Converse to Theorem~\ref{thm:main})}\label{sec:converse:both}
Before we embark on proving the converse, notice that there is no obvious physical degradation in our GBC model 
with noisy feedback shown in Fig.~\ref{fig:bc:one}. In addition, Theorem~\ref{thm:not_useful} cannot be applied as
such.
Interestingly, under the condition in \eqref{eq:thm:main}, an outerbound to the capacity region can be constructed
by equipping each receiver with suitable information about the noise processes in the model. That the proposed outerbound
turns out achievable even without feedback clinches the deal, further underlining the novelty of the
proposed bound. While some steps in our proof 
effectively employ the  degradation between different variables, these are somewhat implicit in the manipulations. 
To make the exposition gradual, we consider a related problem first, that of feedback
only from the strong receiver, where Theorem~\ref{thm:not_useful} turns out useful.

\subsection{Noisy Feedback from Strong Receiver Alone} \label{sec:strong:fb}

\begin{figure}[H]
	\begin{center}
	\subfloat[Original BC~\label{fig:or:strong}]{
	\begin{tikzpicture}[line width=1.0pt]
	
	\node (n0) at (0,0) {$X$};
	\node (r1) at (4,2) {$Y_1$};
	\node (r2) at (4,-1) {$Y_2$};
	\draw[->] (n0) -- (r1);
	\draw[->] (n0) -- (r2);
	
\node (cf) at (2,2.5) {};
	\node (zf) at (cf)[fill=white]  {$+$};
	\draw[<-] (zf) --++(-0,0.5) node[above]{$Z_{\fb1}$};
	\draw[fill=white] (cf) circle(0.25cm) node (nf) {$+$};
	\draw[dashed, ->] (r1) to[bend right=15] (nf);
	\draw[dashed, ->] (nf) to[bend right=35] (n0);
	\end{tikzpicture}} \hspace*{1.0cm}
\subfloat[Augmented BC~\label{fig:augmented}]{
	\begin{tikzpicture}[line width=1.0pt]
	
	\node (n0) at (0,0) {$X$};
		\node (r1) at (4,2)[right] {$Y_1+\Zfa$};
		\node (rr1) at (4,1.1)[right] {${Y}_b$};
		\node (r2) at (4,-1)[right] {$Y_2$};
	\draw[->] (n0) -- (r1.180);
		\draw[->] (n0) -- (rr1);
	\draw[->] (n0) -- (r2);
	
\node (cf) at (2,2.5) {};
	\draw[dashed, ->] (r1.north west) to[bend right=50] (n0);

	\end{tikzpicture}} \hspace*{1.0cm}

%
%

	\caption{Original and Augmented  BCs}
	\end{center}
	\end{figure}
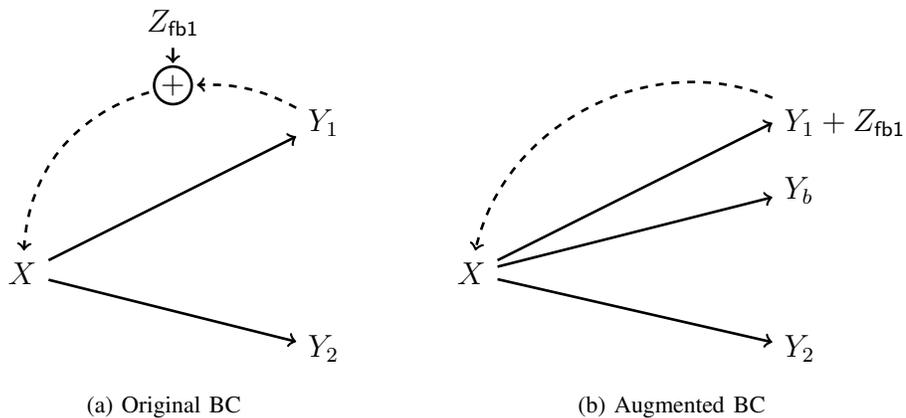

With feedback only from the stronger receiver in a GBC, let us devise the converse for the 
second part of Corollary~\ref{cor:Onesided}. The model is depicted in Fig.~\ref{fig:or:strong}.
We have to show that for $\sftwo=\infty$ and
\begin{equation}\label{eq:condition}
\frac \sone \sfone \leq \frac \stwo \sone -1,
\end{equation}
feedback from the stronger receiver (receiver~$1$) does not enlarge capacity. 



Let us first construct an augmented BC in which receiver~$1$ observes $(Y_1^n+\Zfa^n,Y_b^n)$, where
\begin{equation}\label{eq:tilde}
{Y}_b^n:=Y_1^n- \alpha Z_{\fb1}^n = X^n +\underbrace{ Z_{1}^n -\alpha Z_{\fb1}^n}_{\triangleq {Z}_b^n},
\end{equation} for
\begin{equation}\alpha \triangleq \frac{\sone}{\sfb}. \label{eq:alpha:strong:fb}
\end{equation}
As before, the transmitter observes the feedback signal $Y_{1,i}+Z_{\fb1,i}$ after each channel use $i$. The capacity region $\Cwnfb_{\textnormal{aug}}$ of this augmented BC, which is depicted in Fig.~\ref{fig:augmented},  includes the capacity region $\Cwnfb$ of our original BC:
\begin{equation}\label{eq:line1}
  \Cwnfb \subseteq  \Caug.
\end{equation}
This is immediate since $(1 + \alpha)Y_1 = \alpha(Y_1+\Zfa) + Y_b$, and thus receiver 1 can compute $Y_1^n$ from $Y_1^n+\Zfa^n$ and $Y_b^n$.
We next argue that the augmented BC satisfies the assumptions in Theorem~\ref{thm:not_useful} and thus feedback does not enlarge its capacity region. 
To  this end, 
specialize Theorem~\ref{thm:not_useful} by identifying $\Yta = Y_1+\Zfa$, $\Yha = Y_b$, $\Ytb=\emptyset$, and $\Yhb = Y_2$,  
and  then choose  $P^{\prime}_{\Yhb|\Yha,\Yta}$ such that under this new law
\begin{equation}
    \Yhb = \Yha + \hat{Z}_2,
\end{equation}
with  $\hat{Z}_2$ a zero-mean Gaussian random variable of variance $\stwo - \alpha^2 \sfone- \sone>0$,
 and independent of all other random variables. 
Clearly, \eqref{eq:a1} and \eqref{eq:a3} are immediate under this choice.  The choice of 
parameter $\alpha$ in \eqref{eq:alpha:strong:fb} ensures that the condition 
\eqref{eq:a2} is also met. To see this, notice that
$Z_2$ is independent of $Z_1+\Zfa$ under $P$, whereas $Z_b + \hat Z_2$ is independent of $Z_1+\Zfa$ under $P^{\prime}$. 
Thus, the noise sequences $Z_2^n$ and  $Z_b^n+\hat Z_2^n$ are both independent of the feedback noise
$(Z_1^n+Z_{\fb1}^n)$ in the respective models. Also, notice that the random variables
$Z_2$ and $Z_b + \hat Z_2$ are identically
distributed.  Since all the required assumptions are met, 
we can  employ Theorem~\ref{thm:not_useful} to conclude that feedback does not increase the capacity of the augmented BC, and thus \begin{equation}\label{eq:line3}
\Caug= \Cnofb.
\end{equation}

The capacity region without feedback of the augmented BC is obtained from 
the  well-known  capacity region of a Gaussian vector BC~\cite{GamalKim11}. 
%
Combining \eqref{eq:line1} and \eqref{eq:line3} establishes  the desired converse.


\subsection{Noisy Feedback from both Receivers}
 Observe that feeding back the signal $Y_2$ will allow the transmissions to depend on 
$Z_2+\Zfb$. Thus, there is no obvious stochastic degradation between $Y_2$ and any 
signal derived from the observed symbols at the strong receiver.
Nevertheless, the dependence between $Z_2$ and the transmitted symbols can be decoupled 
by careful conditioning and subsequent manipulations, which enable the identification 
of a suitable degradation structure as shown below. 

By Fano's inequality~\cite{GamalKim11}, after ignoring the $o(n)$ terms,
\begin{align} 
    nR_1 &= I(W_1;Y_1^n|W_2) \nonumber \\
    &\leq I(W_1;Y_1^n,\Zfa^n,Z_2^n +\Zfb^n|W_2) \nonumber \\
    &{=} \sum_{i=1}^{n} I(W_1;Y_{1i}|W_2,Y_1^{i-1},\Zfa^{i-1},Z_2^{i-1} +\Zfb^{i-1}) \nonumber \\
    &\phantom{i}+ I(W_1; {\Zfa}_{,i}, Z_{2i} + 
	{\Zfb}_{,i}|W_2,\Zfa^{i-1},Z_2^{i-1} +\Zfb^{i-1}, Y_1^i) \notag \\
    &\stackrel{(a)}{=} \sum_{i=1}^{n} I(W_1;Y_{1i}|W_2,Y_1^{i-1},\Zfa^{i-1},Z_2^{i-1} +\Zfb^{i-1}) \nonumber \\
    &\stackrel{(b)}{\leq} \sum_{i=1}^{n} h(Y_{1i}|U_i,X_1^{i-1} + Z_{1}^{i-1},\Zfa^{i-1}) - h(Z_{1i}) \nonumber \\
    &\stackrel{(c)}{\leq} \sum_{i=1}^{n} h(Y_{1i}|U_i,X_1^{i-1} + Z_{1}^{i-1}-\gamma \Zfa^{i-1}) - h(Z_{1i}). \label{eq:r1:1}
\end{align}
Here $(a)$ follows since $({\Zfa}_i,Z_{2i}+{\Zfb}_i)$ is independent of $(W_1,W_2,\Zfa^{i-1},Z_2^{i-1} +\Zfb^{i-1},Y_1^{i})$. In $(b)$, we took $U_i=(W_2, Z_2^{i-1}+\Zfb^{i-1})$, and used the fact that $(W_1,U_i,Y_1^{i-1},\Zfa^{i-1})$ determines
the transmitted symbol $X_i$. Clearly, given $X_i$, the remaining uncertainty in $Y_{1i}$ is only due to $Z_{1i}$, which
is independent of $(W_1,U_i,Y_1^{i-1},\Zfa^{i-1})$. The constant in $(c)$ above was taken as $\gamma \triangleq \frac{\sone}{\sfone}$, thus making $(U_i,X^{i-1}+Z_1^{i-1}-\gamma\Zfa^{i-1})$ independent of $Z_{1i}+{\Zfa}_i$, which forms the encoder's information 
about the noise process at receiver~$1$. For the weaker receiver, again by applying Fano's inequality
and ignoring the $o(n)$ terms,
\begin{align}
nR_2
&\leq  h(Y_2^n) - h(Y_2^n|W_2)  \label{eq:bbc:r2:1} \\
&\leq \frac n2 \log 2\pi e (P+\stwo) - h(Y_2^n|W_2). \label{eq:bbc:r2:2}
\end{align}
Let us now expand the second term as
\begin{align}
h(Y_2^n|W_2) &= \sum_{i=1}^{n} h(Y_{2i}|W_2,Y_2^{i-1}) \notag \\
    &\geq \sum_{i=1}^{n} h(Y_{2i}|W_2,X^{i-1} + Z_{2}^{i-1},Z_2^{i-1} + \Zfb^{i-1}) \notag \\
    &=  \sum_{i=1}^{n} h(Y_{2i}|U_i,X^{i-1}\!\!+ Z_{2}^{i-1} \!\! - \beta(Z_2^{i-1} + \Zfb^{i-1})) \notag\\
    &= \sum_{i=1}^{n} h(Y_{2i}|U_i,X^{i-1} + \Tilde{Z}_2^{i-1}), \label{eq:r2:sl}
\end{align}
where  $\beta \triangleq \frac{\stwo}{\stwo + \sftwo}$,
and $\Tilde{Z}_2 \triangleq \frac{\sftwo}{\sftwo + \stwo}Z_2 - \frac{\stwo}{\sftwo + \stwo}Z_{\text{fb}2}$ is independent of $Z_2 + Z_{\text{fb}2}$, the latter being the encoder's information about the noise process at receiver~2.  
Observe  that since  $(\Tilde Z_2, Z_1 - \gamma \Zfa)$ is independent of the encoders' feedback
information $(Z_1+\Zfa, Z_2+\Zfb)$, the former
has no effect on the transmitted symbols.  
Therefore, by denoting $\Tilde{Y}_2 = X + \Tilde Z_2$, 
\begin{align} \label{eq:r2:aux}
 h(Y_{1i}|U_i,X^{i-1}\!\! + Z_{1}^{i-1}\!\!-\gamma \Zfa^{i-1}) \leq   h(Y_{1i}|U_i, \Tilde Y_2^{i-1})
\end{align}
remains true using data processing theorem, as long as
\begin{align} \label{eq:var:order}
    \textnormal{Var}(Z_1 - \gamma Z_{\text{fb}1}) \leq \textnormal{Var}(\Tilde{Z}_2).
\end{align}
%
Notice that \eqref{eq:var:order} is equivalent to
\begin{align} \label{eq:cond:fb:nfb}
    \frac{\stwo}{\sone} \geq \left(\frac{\sone}{\sfone} + 1\right)\left(\frac{\stwo}{\sftwo} + 1\right).
\end{align}
Under the above condition, \eqref{eq:r1:1} yields
\begin{align}\label{eq:r1:sl}
nR_1  \leq \sum_{i=1}^n h(Y_{1i}|U_i,V_i) - h(Z_{1i}),
\end{align}
where we defined $V_i$ as $\Tilde Y_2^{i-1}$.  Also, from \eqref{eq:bbc:r2:1}, 
\begin{align}\label{eq:r2:sl:2}
nR_2 \leq \sum_{i=1}^n h(Y_{2i}) - h(Y_{2i}|U_i,V_i).    
\end{align}
Notice that $(U_i,V_i)\rightarrow X_i \rightarrow (Y_{1i},Y_{2i})$, and we can now obtain single letter rate expressions using one auxiliary random variable $\tilde U = (U,V)$. The optimality of Gaussian auxiliary variable can then be proved along the lines of \cite{GengNair}, by incorporating feedback as in \cite{Viswanathan}. However, we proceed through a more standard route,  by applying the following version of EPI (similar to \cite{Gamal81b}) to connect
\eqref{eq:r1:sl} and \eqref{eq:r2:sl:2}.
\begin{lemma}\label{lem:epi2}
\begin{align} 
2^{\frac{2}{n}\sum\limits_{i=1}^n h(Y_{2i}|U_i,V_i)} 
\geq 2^{\frac{2}{n}\sum\limits_{i=1}^n h(Y_{1i}|U_i,V_i)}    + 2\pi e (\stwo - \sone). \label{eq:bbc:epi}
\end{align}
\end{lemma}
\begin{IEEEproof}
The proof is presented in Appendix~\ref{sec:app:EPI}. 
\end{IEEEproof}
The remaining part of the proof is more routine. Since $\frac n2 \log 2\pi e \stwo \geq h(Y_2^n|W_2) \geq h(Z_2^n)$,  
we can take
\begin{align} \label{eq:r2:semifinal}
h(Y_2^n|W_2) &= \frac n2 \log \bigl( 2 \pi e (\stwo +\theta P) \bigr) , 
\end{align}
for some $\theta \in [0,1]$. Using \eqref{eq:bbc:epi} and the fact that conditioning reduces entropy,
\begin{align}
\sum_{i=1}^n h(Y_{1i}|U_i,V_i) \leq \frac n2 \log  \bigl( 2 \pi e (\sone + \theta P) \bigr).
\end{align}
Using the above two formulas in  \eqref{eq:bbc:r2:2} and \eqref{eq:r1:sl}, we get 
for some $\theta \in [0,1]$,
\begin{align}
R_2 &\leq \frac{1}{2}\log\left( 1 + \frac{(1-\theta)P}{\theta P +
\stwo}\right) \\
R_1 &\leq \frac{1}{2}\log\left(1+\frac{\theta P}{\sone}\right) .
\end{align}
This completes the proof of the converse part of Theorem~\ref{thm:main}.

\section{Zero-Forcing Achievable Schemes for the GBC with Noisy FB \\(Direct Part to Theorem~\ref{thm:main})} \label{sec:achieve}
\def\hZ{\hat{Z}}
\def\hs{\hat{\sigma}}
In order to complete the proof of Theorem~\ref{thm:main}, we now show that a rate pair outside $\Cnofb$ is achievable if the condition in \eqref{eq:thm:main}
is violated by the given tuple $(\sone, \stwo, \sfone, \sftwo)$ of noise variances.
While several feedback coding schemes are available in literature, the main difficulty
is in having tractable rate expressions which can show the required enlargement. Rate regions incorporating noisy feedback are typically stated in terms of the intersections of several hyperplanes, and are thus difficult to express in suitable functional forms for comparison~\cite{shayevitz2012}, \cite{venkataramanan2013}.  Some simplifications are possible, for example, when $\sigma_1^{2}=\sigma_2^{2}$, the rate region proposed in \cite{venkataramanan2013} is shown to achieve rate-pairs outside $\Cnofb$, by suitable substitution of auxiliary variables, and thereby simplifying the expressions. However
extending this  to find the set of noise variances for which an enlargement becomes possible seems difficult
in general.

\subsection{A Simple Linear-Feedback Coding Scheme} 
We will employ linear feedback coding schemes, where the noise realization from each receiver after an odd numbered transmission instant, perceived through the noisy feedback link, is linearly combined and sent along with new symbols in the very next instant. Thus the transmitted signal only depends on feedback during even channel uses, and the feedback values need to be stored for just one instant at the receiver. The fresh symbols at each instant, which are linearly combined with feedback,  are generated and conveyed to all parties, as in standard random coding arguments \cite{GamalKim11}. Specifically, we construct two independent Gaussian codebooks to convey the two messages $W_1$ and $W_2$ to the respective users. Let $U$ denote the codeword symbols to the first receiver and $V$ denote the codeword symbols to the second receiver. 
In order to convey the symbols $(u_{i},v_{i})$ chosen from the codebooks, the transmitter sends,
\begin{align}
    X_{2i-1} &= u_{i} + v_{i} \\
    X_{2i} &= \sqrt{\alpha}\Bigl[ u_{i}-v_{i}+\beta_{1}(Z_{1,2i-1}+Z_{\text{fb}1,2i-1}) \nonumber \\ 
   & \hspace{2cm}+ \beta_{2}(Z_{2,2i-1}+Z_{\text{fb}2,2i-1})\Bigr] \label{eq:EvenTran}
\end{align}
where are $\alpha,\beta_1$ and $\beta_2$ are appropriate real valued parameters. 
Let $\hZ_{j,i}$ be the MMSE estimate of $Z_{j,2i-1}$ given $Z_{j,2i-1} + Z_{\text{fb}j,2i-1}$, for $j=1,2$. Then,
\begin{align}
    \hZ_{j,i} \triangleq \frac{\sigma_{j}^{2}}{\sigma_{j}^{2} + \sigma_{\text{fb}j}^{2}}\left(Z_{j,2i-1} + Z_{\text{fb}j,2i-1}\right),
\end{align}
which has variance $\hs_j^2 \triangleq \frac{\sigma_{j}^{4}}{\sigma_{j}^{2}+\sigma_{\text{fb}j}^{2}}$. Defining 
\begin{equation}\label{eq:gammaj}
\gamma_{j} \triangleq \beta_j \frac{\sigma_j^2}{\hs_j^2}, \quad j=1,2,
\end{equation}
we can rewrite \eqref{eq:EvenTran} as
\begin{align} \label{eq:achieve:gamma}
    X_{2i} = \sqrt{\alpha} \, (u_{i} - v_{i} + \gamma_{1}\hZ_{1} + \gamma_{2}\hZ_{2}).
\end{align}
To facilitate random coding arguments, let us choose the distribution $p(u,v)$
according to  $(U,V)\sim\mathcal{N}\bigl(0,\diag(\thp P, (1-\thp)P)\bigr)$
for some $\thp \in [0,1]$. This choice of $(U,V)$ ensures that the average power constraint is met over the transmissions  $X_{2i-1}$ at odd time instants. To ensure the same for $X_{2i}$, we choose
\begin{align} \label{eq:achieve:alpha}
    \frac{1}{\alpha} = 1 + \frac{\gamma_{1}^{2}\hs_1^2}{P} + \frac{\gamma_{2}^{2}\hs_2^2}{P}.
\end{align}
The following operations are performed at the two receivers for decoding the respective messages.  

\noindent \textbf{Receiver 1:} In two consecutive instants, the observations are
\begin{align}
    Y_{1,2i-1} &= u_{i} + v_{i} + Z_{1,2i-1}  \\
    Y_{1,2i} &= \sqrt{\alpha}(u_{i} - v_{i} + \gamma_{1}\hZ_{1} + \gamma_{2}\hZ_{2}) +Z_{1,2i}.
\end{align}
A simple \emph{zero forcing}  is achieved by computing
\begin{align} \label{eq:achieve:zf:1}
S_{1,i}\triangleq    Y_{1,2i-1} + \frac{Y_{1,2i}}{\sqrt{\alpha}} = 2u_{i} + \gamma_{1}\hZ_{1} + \gamma_{2}\hZ_{2} + Z_{1,2i-1} + \frac{Z_{1,2i}}{\sqrt{\alpha}}.
\end{align}\\
\textbf{Receiver 2:} Here also we do \emph{zero forcing}.
On observing 
\begin{align}
    Y_{2,2i-1} &= u_{i} + v_{i} + Z_{2,2i-1}  \\
    Y_{2,2i} &= \sqrt{\alpha}(u_{i} - v_{i} + \gamma_{1}\hZ_{1} + \gamma_{2}\hZ_{2}) +Z_{2,2i}
\end{align}
in two consecutive instants, the receiver computes
\begin{align} \label{eq:achieve:zf:2}
   S_{2,i}\triangleq   Y_{2,2i-1} - \frac{Y_{2,2i}}{\sqrt{\alpha}} = 2v_{i} - \gamma_{1}\hZ_{1} - \gamma_{2}\hZ_{2} + Z_{2,2i-1} - \frac{Z_{2,2i}}{\sqrt{\alpha}}.
\end{align}
Each receiver $j\in\{1,2\}$ will attempt to decode its  intended message $W_j$ based on the symbols $S_{j,1},\ldots, S_{j,n}$ where $n$ is the codeword length in each codebook.

\noindent \textbf{Analysis of error of probability:} 
Notice that the zero forcing performed  above creates an equivalent point-to-point channel to each receiver, with no interference from the other user's symbols. However the transmissions take place over a blocklength of $2n$ instants now. Therefore, standard random coding arguments \cite{CovTho91} imply that the rate pair $(\Tilde{R}_{1},\Tilde{R}_{2})$ is achievable, where
\begin{align}
    \Tilde{R}_{1} &=
     \frac{1}{4}\log\left(1 + \frac{4\thp P}{\sigma_{1}^{2}\left(1 + \frac{1}{\alpha}\right) + (\gamma_{1}^{2} + 2\gamma_{1})\hs_1^2 + \gamma_{2}^{2}\hs_2^2} \right) \label{eqn:AchievableRate2Noisy}\\
    \Tilde{R}_{2} &=
     \frac{1}{4}\log\left(1 + \frac{4(1-\thp)P}{\sigma_{2}^{2}\left(1 + \frac{1}{\alpha}\right) + (\gamma_{2}^{2} - 2\gamma_{2})\hs_2^2 + \gamma_{1}^{2}\hs_1^2}\right), \label{eqn:AchievableRate2Noisy2}
\end{align}
with  $\alpha$ given by \eqref{eq:achieve:alpha},  and $(\thp, \gamma_1,\gamma_2)$ being arbitrary real tuples satisfying $\thp \in[0,1]$ and $\gamma_1,\gamma_2\geq 0$.
To exemplify the utility of the proposed scheme, consider the case with $\sone=\stwo$, and $\gamma_2=0$, i.e., no feedback from the second receiver. Then, taking $\thp=0.5$ and $\gamma_1=0$ will recover the equal rate point on the no-feedback capacity region. However, some minimal algebra suffices to show that  small negative values of $\gamma_1$ will cause the region given by
~\eqref{eqn:AchievableRate2Noisy} to include rate-pairs outside the no-feedback capacity region. Therefore, this 
scheme almost immediately suggests a capacity enlargement using  passive noisy feedback in a GBC with $\sone=\stwo$. 
Notice the remarkable simplicity when compared to the schemes in \cite{bhaskaran2008}, \cite{venkataramanan2013}, 
however, the latter ones can achieve superior rate-regions.

Generalizing the above idea to different noise variances at the receivers needs more analytical effort. To keep things tractable, we first show that one can obtain rate points outside  $\Cnofb$ at low  average transmit powers, also known as the wideband regime. 

\subsection{Improving on $\Cnofb$ at Low Powers}

Recall our assumption that $\sone \leq \stwo$, and notice that the Pareto optimal rate-pairs on the  boundary of $\Cnofb$ given in \eqref{eq:nofb} can be indexed by the continuous parameter $\theta \in [0,1]$. We fix a
suitable $\theta\in[0,1]$ and show  that for small enough $P$ the corresponding Pareto optimal rate pair in \eqref{eq:nofb} is  dominated by the achievable rate pair $(\tilde R_1, \tilde R_2)$ in  \eqref{eqn:AchievableRate2Noisy} for some
appropriate choice of  $(\thp,\gamma_1, \gamma_2)$.

Consider an arbitrary quadruple $(\theta,\thp,\gamma_1, \gamma_2)$ and define  
\begin{equation}
\zeta\triangleq\gamma_1^2 \hs_1^2 + \gamma_2^2 \hs_2^2.
\end{equation}
By \eqref{eqn:AchievableRate2Noisy}, the pair  $(\Tilde{R}_{1},\Tilde{R}_{2})$ (for the parameters  $\theta^{\prime},\gamma_1, \gamma_2$) dominates the pair $(R_{1},R_{2})$ (for $\theta$) if
\begin{align}
     1 + \frac{4\thp P}{\sigma_{1}^{2}\left(1 + \frac{1}{\alpha}\right)  + 2\gamma_{1} \hs_1^2 + \zeta}
     &\geq \left(1 + \frac{\theta P}{\sone}\right)^{2} \\
     1 + \frac{4(1-\thp)P}{\sigma_{2}^{2}\left(1 + \frac{1}{\alpha}\right) - 2\gamma_{2} \hs_2^2 + \zeta} 
     &> \left(1 + \frac{(1-\theta)P}{\theta P + \stwo}\right)^2,
\end{align}
or equivalently,
\begin{align}
 \frac{4\thp}{\sigma_{1}^{2}\left(1 + \frac{1}{\alpha}\right) + 2\gamma_{1} \hs_1^2  + \zeta} 
 &\geq \frac{2\theta}{\sone} + \frac{\theta^2 P}{\sigma_{1}^4} \label{eq:FinalNonTransformed2Noisy} \\
 \frac{4(1-\thp)}{\sigma_{2}^{2}\left(1 + \frac{1}{\alpha}\right) -  2\gamma_{2})\hs_2^2 + \zeta} 
 &> \frac{2(1-\theta)}{\theta P + \stwo}  +\frac{(1-\theta)^2 P}{(\theta P + \stwo)^2}. \label{eq:FinalNonTransformed2Noisy:2}
\end{align}
Let us change the variables  from  $(\gamma_1, \gamma_2, \thp)$ to $(a_1, a_2, \mu)$ by defining
\begin{align}
    a_j &\triangleq (-1)^j  \frac{\gamma_j\hs_j}{P},\, j =1,2   \label{eq:achieve:transform:1}\\
	\mu &\triangleq \frac 1P \left(\frac{\thp}{\theta} - 1\right). \label{eq:achieve:transform:2}
\end{align}
While there are no restrictions on the parameters $(a_1, a_2)$, the parameter $\mu$ needs to lie in the interval $[-\frac 1P, \frac 1P (\frac 1\theta-1)]$ so that $0 \leq \thp \leq 1$. Notice that the described interval
for possible $\mu$ can be made to include any desired real value by choosing $P$ sufficiently small.

Using the transformations \eqref{eq:achieve:transform:1} --  \eqref{eq:achieve:transform:2} on
\eqref{eq:FinalNonTransformed2Noisy} -- \eqref{eq:FinalNonTransformed2Noisy:2}, we get
\begin{align}
    \frac{4(1+\mu P)}{(a_1^2+a_2^2 )(P^2+P\sone) + 2\sone-2a_1\hs_1 P} 
    &\geq \frac{2}{\sone} + \frac{\theta P}{\sigma_{1}^4}  \\
    \frac{4(1 - \theta(1+ \mu P))}{(a_1^2+a_2^2)(P^2+P\stwo) + 2\stwo-2 a_2\hs_2 P} 
    &> \frac{2(1-\theta)}{\theta P + \stwo} 
	\phantom{w}+ \frac{(1-\theta)^2 P}{(\theta P + \stwo)^2}.
\end{align}
Clearly, the above expressions hold with equality at $P = 0$. Therefore, if the pair of derivatives on the LHS dominates the corresponding RHS derivatives as $P \xrightarrow{} 0^+$, we are done, as this shows that the required capacity enlargement is possible at low enough powers. Differentiating with respect to $P$ and setting $P=0$, we get the conditions
\begin{align}
    2\mu\sone &\geq \theta + (a_1^2 + a_2^2 )\sone - 2a_1\hs_1 
    \\
    (1-\theta)(2a_2 \hs_2 - (a_1^2+a_2^2)\stwo) 
   &> (1-\theta)(1-3\theta) + 2\theta\mu\stwo. 
\end{align}
The first of the above two equations can be guaranteed by choosing
\begin{align} \label{eq:ach:mu}
    \mu \triangleq \frac{\theta}{2\sone} + \frac{a_1^2 + a_2^2 }{2} - \frac{a_1 \hs_1}{\sone},
\end{align}
whereas the second equation, on substitution of \eqref{eq:ach:mu}, will yield 
\begin{align}
   a_1^2&+a_2^2 
	<  \frac{2a_1\theta}{\sqrt{\sone + \sfone}} + \frac{2a_2(1-\theta)}{\sqrt{\stwo + \sftwo}}
   - \frac{\theta^2}{\sone} - \frac{(1-\theta)(1-3\theta)}{\stwo}. \label{eq:ach:sec:cond:2}
\end{align}
Notice that above choice of $\mu$ does not depend on $P$ and thus lies in the desired interval for  all sufficiently small values of $P>0$.
Since $a_1,a_2$ are free parameters, we can choose them as
\begin{align*}
    a_1  &= \frac{\theta}{\sqrt{\sone+\sfone}}\\
     a_2  &= \frac{1-\theta}{\sqrt{\stwo + \sftwo}} . \label{eq:AppendixKKTa}
\end{align*}
By substituting this into \eqref{eq:ach:sec:cond:2}, we need to verify
\begin{align}
    \frac{\theta^2}{\sone} + \frac{(1-\theta)(1-3\theta)}{\stwo} -\frac{\theta^2}{\sone + \sfone} - \frac{(1-\theta)^2}{\stwo + \sftwo} < 0.
\end{align}
for some value of $\theta \in [0,1]$. Defining $x=\frac{\theta}{1-\theta}$, one can equivalently check
if $g(x) < 0$ for some value of $x > 0$, where
\begin{align*}
    g(x) \triangleq &\frac{x\,\sfone}{\sone(\sone + \sfone)} + \frac{\sftwo}{x \, \stwo(\stwo + \sftwo)}   - \frac{2}{\stwo}
     \label{eq:FinalOptimization}.
\end{align*}
In fact, the function  $g(x)$ is minimized for $x>0$ by
\begin{align}
    x^* = \sqrt{\left(\frac{1 + \frac{\sone}{\sfone} }{ 1 + \frac{\stwo}{\sftwo}}\right)\left(\frac{\sone}{\stwo}\right)}.
\end{align}
Now the condition for $g(x^*)<0$ can be seen to be equivalent to
\begin{align}\label{eq:AchievabilityTheorem}
    \frac{\stwo}{\sone} < \left(\frac{\sone}{\sfone}+1\right)\left(\frac{\stwo}{\sftwo} + 1\right),
\end{align}
which is the complement of condition \eqref{eq:thm:main}  given in Theorem~\ref{thm:main}.

Putting it all together, we have shown a capacity enlargement at sufficiently small powers, when at least one of the passive feedback links is not too noisy, as implied by \eqref{eq:AchievabilityTheorem}.

\subsection{Improving on $\Cnofb$ at  All Powers}
We now show that any capacity enlargement at low powers naturally extends  to a capacity enlargement at arbitrary transmit powers. This can be shown by message splitting and bootstrapping the low power achievable scheme.
Consider a rate-pair in which user-$1$ demands a small enough positive rate $R_1$.
Let us split the message $W_{2}$ for user~$2$ into $2$ sub-messages $W_{2,1}$ and $W_{2,2}$ of respective rates 
$R_{2,1}$ and $R_{2,2}$. Consider a small positive $\epsilon$,  and  appropriate parameters $\thp \in[0,1]$ and $\gamma_1,\gamma_2>0$. The message  $W_{2,2}$ is  conveyed to receiver~$2$ using 
a simple point-to-point scheme of power $P-\epsilon$.
We use the variables 
$U \sim \mathcal{N}(0,\thp \epsilon)$ and   $V_{1} \sim \mathcal{N}(0,(1-\thp)\epsilon)$
to denote the code  symbols employed in  the above linear feedback code construction
to encode $(W_1, W_{2,1})$, whereas the variable $V_{2} \sim \mathcal{N}(0, P-\epsilon)$ denotes
the symbol employed in the point-to-point code encoding $W_{2,2}$. All codebooks are generated
independently, and we employ standard random coding arguments to find the error probability.
Over two successive channel uses, the sender then transmits
\begin{align} 
\begin{split}
    X_{2i-1} &= u_{i} + v_{1,i} + v_{2,2i-1} \\
    X_{2i} &= \sqrt{\alpha}(u_{i} - v_{1,i} + \gamma_1 \hZ_{1,2i-1} + \gamma_2 \hZ_{2,2i-1}) + v_{2,2i}.
    \end{split}
\end{align}
Similar to \eqref{eq:achieve:gamma} and \eqref{eq:achieve:alpha}, 
the parameters $\alpha, \gamma_1$, and $\gamma_2$ are chosen so that the sum $ \sqrt{\alpha}(u_{i} - v_{1,i} + \gamma_1 \hZ_{1,2i-1} + \gamma_2 \hZ_{2,2i-1})$ satisfies 
the average power constraint of $\epsilon$.

Receiver 2 infers its desired messages via the following steps:
\begin{itemize}
    \item It decodes message $W_{2,2}$ by treating everything as noise.
    Clearly the transmissions in successive instants are independent, and we can treat the rest of the transmissions as memoryless  Gaussian noise sequences
    while decoding the $V_{2}$ codewords.
    \item After inferring $W_{2,2}$, it subtracts the $V_{2}$ codeword to obtain a \emph{more clean}
	 BC with equivalent average transmit power $\epsilon$, 
		and it applies the \textit{zero forcing} decoding scheme as in  \eqref{eq:achieve:zf:2}.
\end{itemize}
Receiver~$1$ decodes in a similar way: it first decodes messages $W_{2,2}$ (even though this message is not intended for it), subtracts the  $V_{2}$ codeword, and applies the zero forcing decoding scheme  described in \eqref{eq:achieve:zf:1}.

We show that for appropriate choices of $\epsilon, \thp, \gamma_1,\gamma_2$, the proposed scheme achieves a rate-point outside the no-feedback capacity region $\Cnofb$. We first notice that  $R_1$ and $R_{2,1}$ can be chosen as the rates $\tilde{R}_1$ and $\tilde{R}_2$ in \eqref{eqn:AchievableRate2Noisy} and  \eqref{eqn:AchievableRate2Noisy2},   however  with the average power $P$ replaced by $\epsilon$. We have shown in the previous subsection that for sufficiently small $\epsilon>0$ there exist choices of $\thp, \gamma_1, \gamma_2$ and $\theta$ such that 
	\begin{align}
	&{R}_{1} = \frac{1}{2}\log_{2}\left(1 + \frac{\theta\epsilon}{\sigma_{1}^{2}}\right),\\
	& {R}_{2,1}  > \frac{1}{2}\log_{2}\left(1 + \frac{(1-\theta)\epsilon}{\theta\epsilon+\sigma_{2}^{2}}\right).
	\end{align}
	We continue with such a sufficiently small value of $\epsilon$. 
	Since Message $W_{2,2}$ can be transmitted at a rate 
	\begin{align}
	R_{2,2}= \frac{1}{2} \log_2 \left( 1 + \frac{P-\epsilon}{\epsilon+\sigma_2^2}\right), 
	\end{align}
	by introducing $\thpp \triangleq \theta \frac{\epsilon}{P}$, we conclude that the rate pair 
\begin{align} 
R_1 & = \frac{1}{2}\log_{2}\left(1 + \frac{\thpp P}{\sigma_{1}^{2}}\right),\\
     R_2& = {R}_{2,1} + R_{2,2} \nonumber\\
   & > \frac{1}{2}\log_{2}\left(1 + \frac{(1-\theta)\epsilon}{\theta\epsilon+\sigma_{2}^{2}}\right) + \frac{1}{2}\log_{2}\left(1 + \frac{P-\epsilon}{\epsilon + \sigma_{2}^{2}}\right) = \frac{1}{2}\log_{2}\left(1 + \frac{(1 - \thpp)P}{\thpp P + \sigma_{2}^{2}}\right).
\end{align}
is achievable. Observing that  this rate-pair  lies outside $\Cnofb$ is sufficient to conclude the proof.
In short, we used the enlargement for small values of power, and allotted all the remaining power for  transmissions to the weak receiver. This allowed us to strictly improve the no-feedback capacity region at high powers as well. 


\section{Conclusion} \label{sec:conc}
We have analyzed the effect of passive noisy feedback in enlarging the capacity region
of a Gaussian broadcast channel. Interestingly, too much noise in both the feedback links 
does not lead to any enlargement in the asymmetric user case, in sharp contrast with the case
of identical channel transition laws to the receivers. For the latter case,  it is known that
any noise of finite variance in the feedback link is beneficial for capacity
enlargement. While we have characterized the regime of 
noise variances for which no enlargement occurs, our achievable 
scheme shows an enlargement outside this proposed regime, thus making the characterization
sharp. While showing an enlargement, however small, was sufficient for our purposes here, 
there is still some way to go for finding the actual capacity region with enlargement. More sophisticated
coding schemes than the ones proposed here may be required to achieve that. 

The genie aided converse constructions, and associated ideas, seem to be beneficial
in analyzing other multiuser systems with feedback. This is currently under consideration.

\appendices
\section{Proof of Lemma~} \label{sec:app:EPI}
\def\tZ{\tilde{Z}}
\begin{IEEEproof}
	The proof is very similar to that in~\cite{Gamal81b}. For the sake of completeness, here we 
repeat the arguments, which
	proceeds by induction on $n$. 
For $n=1$, the inequality follows from
entropy power inequality~\cite[pg. 22]{GamalKim11} since we may write
$$ 
h(Y_{21}|U_1,V_1) = h(Y_{11}+\tZ_1|U_1,V_1),
$$
where $\tZ_1\sim \mathcal N(0,\stwo-\sone)$. Notice that $\tZ_i$ is
independent of the transmitted symbols, messages and other noise processes. 
Similarly, because
$Z_{1m}$ and $Z_{2m}$ are both Gaussian of variances $\Var(Z_{1m})\leq \Var(Z_{2m})$ and independent of $(U_m,V_m,X^m)$, we can write for any integer $m\geq 1$:
\begin{align}
h(Y_{2m}|U_m,V_m) = h(Y_{1m}+\tZ_{m}|U_m,V_m),
\end{align}
for some zero-mean Gaussian 
 $\tZ_m$ of variance $\Var(\tilde{Z}_{m})=\stwo-\sone$  independent of $(Y_{1,m}, U_m, V_m)$.


Now, assume that \eqref{eq:bbc:epi} is true for $n=m-1$. 
By the conditional EPI~\cite{GamalKim11},
\[ 2^{2h(Y_{2m}|U_{m},V_{m})} \geq 2^{2 h(Y_{1m}|U_{m},V_{m})} +
2^{2 h(\tZ_m)}.\]
i.e.,
\begin{align*}
	2h(Y_{2m}|U_{m},V_{m})) &\geq  
	\log\left( 2^{2 h(Y_{1m}|U_{m},V_{m})} +
2\pi e (\stwo-\sone)\right).
\end{align*}
Therefore,
\begin{align*}
\frac{2}{m}\sum_{i=1}^{m}h(Y_{2i}|U_i,V_i)
	&= \frac{m-1}{m}\frac{2}{m-1}\sum_{i=1}^{m-1}h(Y_{2i}|U_i,V_i) + \frac{2}{m}h(Y_{2m}|U_m,V_m)\\
&\stackrel{\text{(a)}}{\geq}
\frac{m-1}{m}\log\left( 
    2^{\frac{2}{m-1} \sum_{i=1}^{m-1} h(Y_{1i}|U_i,V_i)} 
    + 2\pi e (\stwo-\sone) \right)\\
&\qquad+
\frac{1}{m}\log\left(
    2^{2 h(Y_{1m}|U_{m},V_{m}))} + 2\pi e (\stwo-\sone)
\right)\\
&\stackrel{\text{(b)}}{\geq}
\log\left(
2^{\frac{2}{m}\sum_{i=1}^m
h(Y_{1i}|U_{i},V_i)} + 2\pi e (\stwo-\sone)
\right).
\end{align*}
where (a) follows from the induction hypothesis and the EPI above, and (b)
follows from convexity of $\log(2^u+v)$ in $u$ for $v\geq 0$.
\end{IEEEproof}

\ifCLASSOPTIONcaptionsoff
  \newpage
\fi

\bibliographystyle{IEEEtran}
\bibliography{./Ref}




\end{document}